\documentclass[journal,twoside]{IEEEtran}
\usepackage{amsmath}%
\usepackage{graphicx}
\usepackage{setspace} 
\usepackage[justification=centering,font=footnotesize]{caption}
\usepackage{cite}
\usepackage{flushend}
\usepackage[utf8]{inputenc} 

\newcommand{\nn}{\nonumber}
\DeclareMathOperator*{\argmin}{arg\,min}

\ifCLASSINFOpdf
\else
\fi

\begin{document}
\bstctlcite{IEEEexample:BSTcontrol}
%
\title{SVA Based Beamforming}

\author{
        Erdal~Epçaçan~
        and~Tolga~Çiloğlu
\thanks{The authors are with the Department of Electrical and Electronics Engineering,
Middle East Technical University, Ankara, 06800 TURKEY (e-mail:
erdalepcacan@gmail.com; ciltolga@metu.edu.tr)}
}

\markboth{}%
{ Epçaçan and Çİloğlu: SVA Based Beamforming}





\maketitle

\begin{abstract}
A frequency domain method is proposed to reduce the sidelobe level of a uniformly weighted uniform linear array in direction-of-arrival estimation. The development is based on the nonlinear method of spatially variant apodization originally proposed for spectral analysis and synthetic aperture radar imagery.
    
\end{abstract}

\begin{IEEEkeywords}
SVA, beamforming, sidelobe suppression
\end{IEEEkeywords}

%

\section{Introduction} \label{intSTBF}

Short-time Fourier transform is a basic tool in spectral analysis. In short-time Fourier transform, “short” data/signal segments are generally multiplied by a weighting function (window) prior to the Fourier transform computation. This operation is known as windowing or apodization, the latter term being more widely used in optics. A short-time Fourier spectrum involves the convolution of the Fourier transforms of the signal and the window. Broadly speaking, the effect of the Fourier transform of the window (on the resulting short-time spectrum) is generally quantified in terms of its main lobe width and side lobe level.  Main lobe width is related to the accuracy in localizing spectral peaks and side lobe level to the interference among spectral components. Inherent tradeoff between main lobe width and side lobe level has led to a substantial amount of study on the design of window functions \cite{onWindows, desOfEffWind, aCompStdOnWin} ad references therein. On the other hand in \cite{nonApod}, it has been shown that by applying a nonlinear operation, the sidelobe level can be reduced without degrading the mainlobe resolution. The idea is to first compute multiple STFTs each with its own window function and then, for each frequency, keep the value of the one having the minimum magnitude. Specifically, let the set of windows be $w_i[n]$,  input data be $x[n]$ and windowed data be $y_i[n] = w_i[n]x[n]$. Then the output, $Y(e^{j\omega})$ for this method is:

 \begin{equation}
Y(e^{j\omega})= Y_i(e^{j\omega}); ~i = \argmin_k |Y_k(e^{j\omega})| 
\label{eq:svaToBF0}
\end{equation}

\noindent Where $Y_i(e^{j\omega})$ is the DTFT of $y_i[n]$.
 
This approach has been utilized in \cite{nonApod} for sidelobe reduction in SAR (synthetic aperture radar) imagery. The use of two (double apodization) or more (multi-apodizaton) windows has been studied. Although, it would be preferred to include as many different types of windows as possible in multi-apodization, this may not be practical because of the accompanying computational demand. In \cite{nonApod}, another method, a special case of multi-apodization, called as Spatially Variant Apodization (SVA) has been proposed to incorporate infinitely many windows provided that windows are of raised-cosine type. In \cite{svaFourRec,effSVA}, SVA has been formulated as a spectral estimator and its relationship to minimum variance spectral estimator (MVSE) has been emphasized. Although SVA is data dependent, it effectively suppresses the sidelobe level without explicit use of a priori information. Moreover, it has much less computational demand compared to MVSE and free from its finite numerical problems since it does not require the inversion of the covariance matrix of the input. 

Although, in \cite{nonApod} it has been stated that nonlinear apodization techniques have very broad range of applications where the data can be represented as the Fourier transform of a finite-aperture signal and there are other studies on SVA \cite{gsva, rsva, applySVA}, any study adapting SVA to beamforming has not been encountered. As in other signal processing operations, narrow main lobe and low side lobe level in spatial power spectrum is very important for beamforming and direction of arrival (DOA) estimation. In this paper, an SVA approach will be developed for spatial domain processing of the element outputs of a sensor array. To achieve the goal we use the similarity of Fourier transform and beamforming operations. In this development beamforming is performed in frequency domain. 

In Section \ref{svaSTBF} we review SVA. In Section \ref{fftBf} the relationship between DTFT and beamforming is revealed. SVA based beamforming is presented in Section \ref{algSTBF}. Section \ref{resSTBF} provides some simulation results. Some of the practical issues are discussed in Section \ref{prcSTBF}. Conclusions are given in Section \ref{concSTBF}. 

\IEEEpubidadjcol

\section{Spatially Variant Apodization (SVA)} \label{svaSTBF}
SVA uses an infinite set of windows which are all in the form of a raised-cosine function. The family of raised-cosine window functions are given by 

\begin{equation}
w(n;\alpha)=1-2\alpha cos(2\pi n/N), n=0,1,\ldots,N-1
\label{eq:svaToBF1}
\end{equation}

\noindent where $N$ is the window length and  
\begin{equation}
0 \leq \alpha  \leq 1/2.
\label{eq:svaToBF2}
\end{equation}

The extreme values $\alpha=0$ and $\alpha=1/2$ yield rectangular and Hanning windows, respectively. The DTFT of the raised-cosine function given in \eqref{eq:svaToBF1} is

\begin{equation}
W(e^{j\omega}) =2\pi(-\alpha \delta(e^{j(\omega-2\pi/N)})+\delta(e^{j\omega})-\alpha \delta(e^{j(\omega+2\pi/N)})).
\label{eq:svaToBF3}
\end{equation}

As the DTFT of the window function \eqref{eq:svaToBF3} contains only three impulses (Dirac delta functions), windowing can be performed easily in frequency domain.  Let $X(e^{j\omega})$ be the DTFT of the observation then the DTFT of the windowed observation is

\begin{align}
Y(e^{j\omega})&=-\alpha X(e^{j(\omega-2\pi/N)})+X(e^{j\omega})\nn \\
&-\alpha X(e^{j(\omega+2\pi/N)}).
\label{eq:svaToBF4}
\end{align}

For each frequency, SVA finds the value of $\alpha$ which minimizes $|Y(e^{j\omega})|^2$ subject to the constraint $0 \leq \alpha \leq 1/2$. The problem is solved by setting partial derivative of $|Y(e^{j\omega})|^2$ with respect to $\alpha$ to zero and solving for $\alpha$. The optimal value is  \cite{nonApod}

\begin{equation}
\alpha_0(\omega)=Re \bigg\{\frac{X(e^{j\omega})}{X(e^{j(\omega-2\pi/N)})+X(e^{j(\omega+2\pi/N)}} \bigg\}.
\label{eq:svaToBF5}
\end{equation}

Therefore, the SVA output with the constraint in \eqref{eq:svaToBF2} becomes:

 \begin{equation}
Y(e^{j\omega}) = \begin{cases}
				X(e^{j\omega})  & \alpha_0(\omega)<0,\\
				X(e^{j\omega})-\alpha_0(\omega) S(e^{j\omega}) & 0 \leq \alpha_0(\omega) \leq 1/2,\\
				X(e^{j\omega})-1/2S(e^{j\omega}) &\alpha_0(\omega)> 1/2
			\end{cases}
\label{eq:svaToBF6}
\end{equation}

\noindent where $S(e^{j\omega})=(X(e^{j(\omega-2\pi/N)})+X(e^{j(\omega+2\pi/N)}))$. The result in \eqref{eq:svaToBF6} is obtained by processing the real and imaginary parts of $X(e^{j\omega}) $ jointly and is called "I-Q jointly SVA", where I and Q are in-phase and quadrature components of the complex data respectively. In \cite{nonApod}, a different approach that processes the real and the imaginary parts separately is also given  and is called "I-Q separately SVA". The output for "I-Q separately SVA" is given as

 \begin{equation}
Y(e^{j\omega}) = \begin{cases}
				X(e^{j\omega})  & \alpha_0(\omega)<0,\\
				0& 0 \leq \alpha_0(\omega) \leq 1/2,\\
				X(e^{j\omega})-1/2S(e^{j\omega}) &\alpha_0(\omega)> 1/2.
			\end{cases}
\label{eq:svaToBF7}
\end{equation}

This is applied to both in-phase and quadrature components of the complex data.

In practice DTFT will be implemented by DFT, the expressions for "I-Q jointly SVA" and "I-Q separately SVA" respectively in terms of $N$-point DFT are

 \begin{equation}
Y[k] = \begin{cases}
				X[k]  & \alpha_0[k]<0,\\
				X[k]-\alpha_0[k](S[k])  & 0 \leq \alpha_0[k] \leq 1/2,\\
				X[k]-1/2S[k] &\alpha_0[k]> 1/2
			\end{cases}
\label{eq:svaToBF8}
\end{equation}

 \begin{equation}
Y[k] = \begin{cases}
				X[k]  & \alpha_0[k]<0,\\
				0  & 0 \leq \alpha_0[k] \leq 1/2,\\
				X[k]-1/2S[k] &\alpha_0[k]> 1/2
			\end{cases}
\label{eq:svaToBF9}
\end{equation}

\noindent where $S[k] = X[k-K]+X[k+K]$, $k = 0,1,...,N-1$, $K$ is zero padding factor and $\alpha_0[k]$ is given as:

\begin{equation}
\alpha_0[k]=Re \bigg\{\frac{X[k]}{X[k-K]+X[k+K]} \bigg\}.
\label{eq:svaToBF10}
\end{equation}

\begin{figure} %
\centering
\includegraphics[width=1\columnwidth]{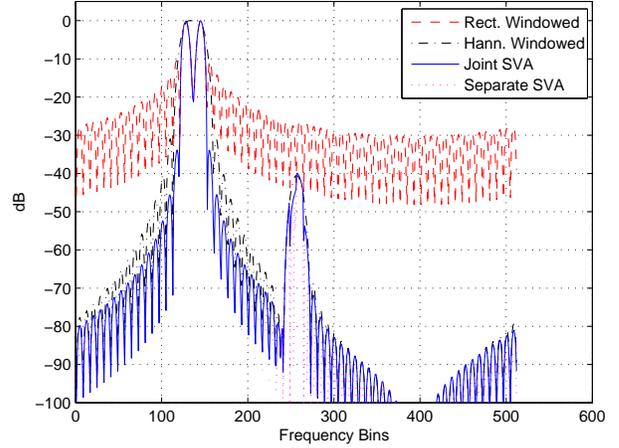}%
\caption{SVA for two close, equal power  and one distant low power sources}%
\label{fig:svaToBF1}%
\end{figure}

An example for "I-Q jointly SVA" and "I-Q separately SVA" is provided in Figure \ref{fig:svaToBF1}. It is seen that, SVA suppresses sidelobes while preserving the mainlobe resolution. In this technique for each frequency an optimal $\alpha_0 (k)$ value is found depending on the data to be processed.

\section{Relationship between DTFT and Uniform Linear Array Beamformer} \label{fftBf}

The conventional beamformer output of a complex plane wave of frequency $\omega_0$ and DOA of $\phi$ for a uniform linear array (ULA) of $M$ sensors is \cite{arSigProCT}

\begin{equation}
X_{BF}(\phi)=\sum_{m=1}^M{x_m e^{-j\omega_0 \tau_m (\phi)}} 					
\label{eq:svaToBF11}
\end{equation}

\noindent where $\phi$ is defined with respect to positive $x$-axis and sensors are located on $x$-axis, $x_m$ is the spectral component of the $m^{th}$ sensor at frequency $\omega_0$ and $\tau_m (\phi)$ is the time delay at the $m^{th}$ sensor with respect to reference point. Assuming the reference point as the first sensor of the array $\tau_m (\phi)$ is defined as

\begin{equation}
\tau_m (\phi) = \frac{(m-1)d\cos\phi}{c}
\label{eq:svaToBF12}
\end{equation} 

\noindent where $d$ is the separation between the sensors and $c$ is the propagation speed of the wave. Letting $d = d_1 \lambda_0,~ d_1>0$ where $\lambda_0 = 2\pi c/\omega_0$,  \eqref{eq:svaToBF11} can be written as

\begin{align}
X_{BF}(\phi) &=\sum_{m=0}^{M-1}{x_m e^{-j\omega_0 \frac{md}{c}\cos(\phi)}} \nn \\
 &= \sum_{m=0}^{M-1}{x_m e^{-jm2 \pi d_1 \cos(\phi)}}. 		
\label{eq:svaToBF13}
\end{align}

On the other hand, DTFT of a length-$M$ signal $x[m]$ is

\begin{equation}
X(e^{j\omega})=\sum_{m=0}^{M-1}{x[m] e^{-j\omega m}}. 					
\label{eq:svaToBF14}
\end{equation}

Comparing equations \eqref{eq:svaToBF13} and \eqref{eq:svaToBF14}, when $x[m]=x_m$, $X_{BF}(\phi)$ and $X(e^{j\omega})$ are related as

\begin{align}
X_{BF}(\phi) &= X(e^{j\omega_0 \frac{d}{c}\cos(\phi)}) = X(e^{j 2\pi d_1\cos(\phi)})\nn, \\
X(e^{j\omega}) &= X_{BF}\bigg(\cos^{-1}\bigg(\frac{\omega c}{\omega_0 d}\bigg)\bigg)\\
								&= X_{BF}\bigg(\cos^{-1}\bigg(\frac{\omega}{2\pi d_1}\bigg)\bigg)\nn.\\
\label{eq:svaToBF15}
\end{align}

Therefore the beamformer output of ULA at direction $\phi$  can be found by evaluating the Fourier transform of the ULA output at frequency $\omega=(\omega_0 d \cos(\phi))/c$.  In other words, given $X(e^{j\omega})$  we can obtain $X_{BF}(\phi)$ by a nonlinear mapping. 

In terms of the DFT values, the results in \eqref{eq:svaToBF15} can be approximated as:

\begin{align}
X_{BF}(\phi) &\approx X[int(N\cos(\phi)d_1)] \nn, \\
X[k] &\approx X_{BF}\bigg(\cos^{-1}\bigg(\frac{k}{d_1N}\bigg)\bigg)
\label{eq:svaToBF16}
\end{align}

\noindent where $X[k]$ is the $N-$point DFT of the ULA output. In \eqref{eq:svaToBF16}, $int(.)$ stands for rounding operation, i.e. the integer closest to the argument.

\section{SVA Based Beamforming }\label{algSTBF}

Beamforming by a ULA and DTFT operations are similar. Using this similarity and the fact that SVA is based on the DTFT of the observation, SVA can be applied to beamforming.

Let the output of the $m^{th}$ sensor in a ULA with $M$ sensors be $\tilde{x}_m[n], m = 0, 1, ... M-1$ and its DTFT be $\tilde{X}_m(e^{j\omega})$. Also let the value of the DTFT at frequency $\omega_0$ be $x_m =\tilde{X}_m(e^{j\omega_0})$ and DTFT of $\textbf{x} = [x_0~ x_1 ~...~ x_{M-1}]$ be $X_s(e^{j\omega})$ ("s" stands for spatial),

\begin{equation}
X_s(e^{j\omega}) = \sum_{m=0}^{M-1}{x_me^{-j\omega m}}.
\label{eq:svaToBF17}
\end{equation}

Recall that "I-Q jointly SVA" output for a given DTFT, $X(e^{j\omega})$, is given by \eqref{eq:svaToBF4} with $\alpha_0(\omega)$ given by \eqref{eq:svaToBF5}

Accordingly, using the relationship in \eqref{eq:svaToBF15}, SVA beamforming output can be found as

\begin{equation}
Y_{BF}(\phi) = Y(e^{j 2\pi d_1\cos(\phi)}). 
\label{eq:svaToBF20}
\end{equation}

Using \eqref{eq:svaToBF4} and \eqref{eq:svaToBF20}:

\begin{align}
Y_{BF}(\phi) &= -\alpha_0(\phi) X_s(e^{j(2\pi d_1\cos(\phi)-2\pi/M)})\nn\\
							+X_s(e^{j2\pi d_1\cos(\phi)})
							&- \alpha_0(\phi) X_s(e^{j(2\pi d_1\cos(\phi)+2\pi/M)}). 
\label{eq:svaToBF21}
\end{align}

Now using the second line of \eqref{eq:svaToBF15}, SVA-beamforming output becomes

\begin{align}
Y_{BF}(\phi) &= -\alpha_0(\phi) X_{BF}\bigg(\cos^{-1}\bigg(\cos(\phi)-\frac{1}{M d_1}\bigg)\bigg)\nn\\
							&+X_{BF}\big(\phi\big) \nn\\ 
							&- \alpha_0(\phi) X_{BF}\bigg(\cos^{-1}\bigg(\cos(\phi)+\frac{1}{M d_1}\bigg)\bigg). 
\label{eq:svaToBF22}
\end{align}

$\alpha_0(\phi)$ is given by

\begin{align}
\alpha_0(\phi)&=Re \bigg\{\frac{X_{BF}\big(\phi\big)}{S_{BF}(\phi)} \bigg\},\nn\\
  ~~~~~~~~&0 \leq \alpha_0(\phi) \leq 0.5
\label{eq:svaToBF23}
\end{align}

where $S_{BF}(\phi)=X_{BF}\bigg(\cos^{-1}\bigg(\cos(\phi)-\frac{1}{M d_1}\bigg)\bigg)+X_{BF}\bigg(\cos^{-1}\bigg(\cos(\phi)+\frac{1}{M d_1}\bigg)\bigg)$

In terms of DFT, SVA beamforming output is given by

\begin{equation}
Y_{BF}(\phi) \approx Y[int(N\cos(\phi)d_1)] 
\label{eq:svaToBF2---}
\end{equation}

\noindent where $Y[k]$ is obtained by applying SVA to $X_s[k]$, (equations \eqref{eq:svaToBF8}-\eqref{eq:svaToBF10}), and $X_s[k]$ is the $N$-point DFT of $\textbf{x} = [x_0~ x_1 ~...~ x_{M-1}]$. Here zero padding factor is $K = N/M$ since $x_m = 0$ for $m\geq M$.

\section{Results}\label{resSTBF}

\begin{figure}[!htb]%
\centering
\includegraphics[width=1\columnwidth]{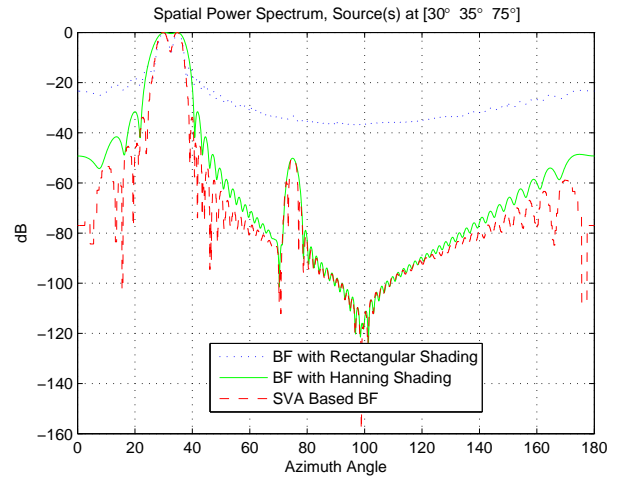}%
\caption{SVA Based Beamforming}%
\label{fig:svaToBF5}%
\end{figure}

Mainlobe widths (resolution) and sidelobe levels, obtained by SVA based beamforming and by alternative methods will be compared. Figure \ref{fig:svaToBF5} shows an example with 3 targets in the environment. The target at azimuth angle of $75^\circ$ has $50~dB$ lower power compared to the other two targets. In this example a ULA with 64 sensors with $d_1 =1/2$ has been used. SNR is infinite and 1024-point DFT is used. It is seen that beamforming with Hanning shading cannot resolve two close targets and beamforming with rectangular shading cannot find the low power target whereas SVA beamforming can resolve two close sources and find the low power target. In the example above, "I-Q jointly SVA" has been applied.

\section{Some Practical Issues} \label{prcSTBF}

\begin{figure}[!htb]%
\centering
\includegraphics[width=1\columnwidth]{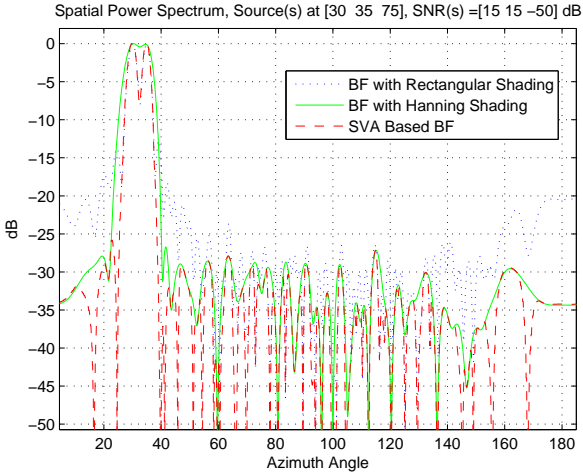}%
\caption{SVA Based Beamforming with noise}%
\label{fig:svaToBF7}%
\end{figure}

Additive noise: The method presented in this paper inherits the properties of SVA therefore it has no effect on noise, that is sidelobe level can be reduced at most to the level of noise. The results of an example with the same parameters of the case given in Figure \ref{fig:svaToBF5} but with additive noise at specified SNRs are provided in Figure \ref{fig:svaToBF7}. Figure \ref{fig:svaToBF7} reveals that for the regions where sidelobe level is greater than the noise level, around mainlobe, the sidelobe has been suppressed to the noise level, however for other regions sidelobe level remains at the noise level.

\begin{figure}[!htb]%
\centering
\includegraphics[width=1\columnwidth]{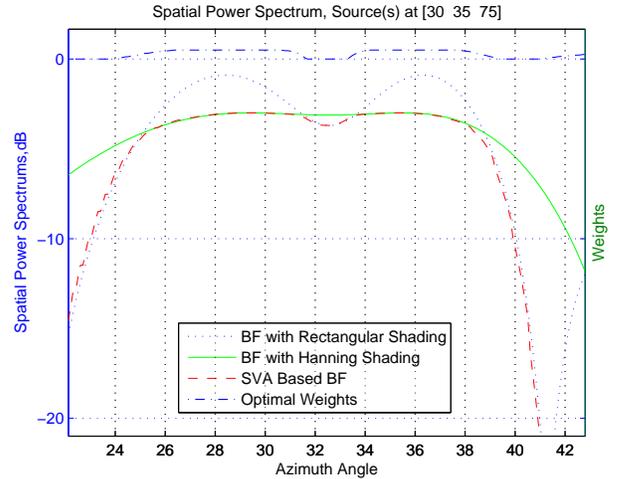}
\caption{SVA Based Beamforming with 32 sensors}%
\label{fig:svaToBF8}%
\end{figure}

\begin{figure}[!htb]%
\centering
\includegraphics[width=1\columnwidth]{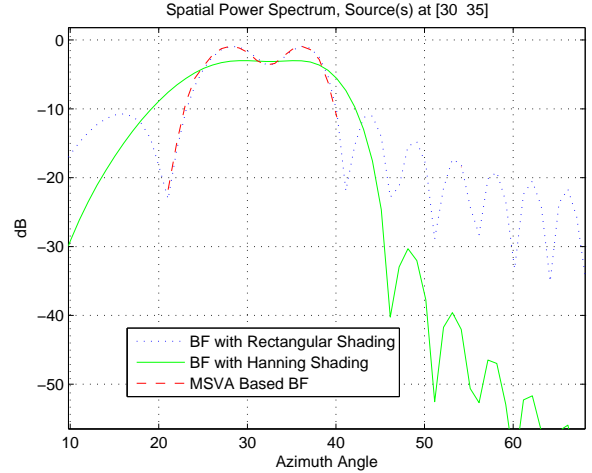}
\caption{MSVA Based Beamforming with 32 sensors}
\label{fig:svaToBF9}
\end{figure}

The peak-to-peak variation around the target peaks: The number of sensors affects the resolving capability, there may be cases where rectangular shading has better resolution compared to that of SVA based beamforming. The results for two close targets, with the same parameters of the case given in Figure \ref{fig:svaToBF5} but with 32 sensors, are provided in Figure \ref{fig:svaToBF8}. As it is seen from Figure \ref{fig:svaToBF8} there is about $2~dB$ loss compared to rectangular shading. The reason for this is the decrease in mainlobe energy \cite{MSVA, superMSVA}. When Figure \ref{fig:svaToBF8} is analyzed it is seen that around the mainlobe the optimal weights are close to $1/2$ instead of $0$. To have the same resolution with rectangular shading, the number of sensors has to be increased or  modified SVA (MSVA) method described in \cite{MSVA, superMSVA} can be applied. The result of MSVA based beamforming is given in Figure \ref{fig:svaToBF9}. It is seen that MSVA and rectangular shading curves almost coincides around the peaks.

The number of DFT points: The number of DFT points has to be large enough to have an acceptable accuracy and better performance. Increasing the number of DFT points decreases the error in nonlinear mapping. Moreover, having the number of DFT points as an integer multiple of the number of sensors has a positive effect on the performance.

\section{Conclusion}\label{concSTBF}

As a specific multi-apodization approach, SVA is a computationally convenient method that brings together the advantages of different window functions in spectral analysis. In this paper SVA has been adapted to beamforming by a uniform linear array. The main idea is to use the similarity of beamforming and DTFT operations and the fact that optimization in SVA is done in frequency domain. SVA based beamforming achieves the resolving capability of rectangular shading with much lower sidelobe level. SVA based beamforming naturally inherits the properties of SVA; A possible drawback, its poorer resolution (relative to that of rectangular shading) when the number of sensors is reduced can be overcome by a modified SVA method. Choosing the number of DFT points as an integer multiple of the number of sensors reduces possible deviation in array output pattern to a negligible level. That this is a frequency domain method can be considered as a practical advantage due to the availability of custom FFT routines for many processors.

\ifCLASSOPTIONcaptionsoff
  \newpage
\fi


\newpage

	\begin{thebibliography}{}
%
%
	 \bibitem{onWindows}
	F. J. Harris,	On the use of windows for harmonic analysis with the Discrete Fourier transform,	Proceedings of IEEE,	51-83 (1978)
	
		\bibitem{svaFourRec}
	Jung Ah C. Lee and David C. Munson(Jr),	Spatially Variant Apodization for Image Reconstruction from Partial Fourier Data,	Image Processing, IEEE Transactions on,	2000,	Nov.,	1914-1925
	 
	
	\bibitem{effSVA}
	Jung Ah C. Lee and David C. Munson(Jr),
	Effectiveness of spatially-variant apodization,
	Image Processing, 1995. Proceedings., International Conference,
	1995,
	Oct.,
	147-150
	 
	\bibitem{desOfEffWind}
  Vivek Kumar and others ,
  Design of effective window function for FIR filters,
  Advances in Engineering and Technology Research (ICAETR), 2014 International Conference on,
  Unnao, India ,
  Aug.,
  2014

\bibitem{aCompStdOnWin}
  Hrishi Rakshit and Muhammad Ahsan Ullah ,
   A comparative study on window functions for designing efficient FIR filter,
  Strategic Technology (IFOST), 2014 9th International Forum on,
  Cox's Bazar,
  Oct.,
 2014,
  91-96


\bibitem{gsva}
  Brian Hendee Smith,
  Generalization of Spatially Variant Apodization to Noninteger Nyquist Sampling Rates,
  IEEE Transaction on Image Processing,
  9,
  6,
  2000,
  1088-1093


\bibitem{rsva}
  Carlos Castillo-Rubio and Sergio Llorente-Romano and Mateo Burgos-Garcia,
  Robust SVA Method for Every Sampling Rate Condition,
  IEEE Transactions on Aerospace and Electronic Systems Vol. 43, No. 2 April 2007,
  43,
  2,
  2007,
  571-580


\bibitem{applySVA}
  W. Zhai, Y. Zhang ,
  Apply Spatially Variant Apodization to SAR/INSAR Image Processing,
   Strategic Technology (IFOST), 2014 9th International Forum on,
  Cox's Bazar,
   Oct.,
  2014,
  91-96

	 \bibitem{nonApod}
	H. C. Stankwitz and R.J. Dallaire and J.R. Fienup,
	Nonlinear apodization for  sidelobe control  in SAR Imagery,
	IEEE Trans. Aerosp. Electron.Syst,
1995,
Jan.,
23-52,
	31,
	 
	\bibitem{arSigProCT}
    Array  Signal Processing, Concept and Techniques,
    D.  H. Johnson  and D. E. Dudgeon.,
    1993,
    Prentice-Hall,
    Englewood Cliffs, NJ,
      
		
		\bibitem{MSVA}
	Ni Chong and others,
	A SAR sidelobe suppression algorithm based on modified spatially variant apodization,
Sci China Tech Sci,
	2010,
	March,
	2542-2551,
	53,


\bibitem{superMSVA}
	Ni Chong and others,
A super-resolution algorithm for synthetic aperture radar based on modified spatially variant apodization,
	 Science China Physics, Mechanics, Astronomy,
	2010,
	Feb.,
	355-364,
	54,


\end{thebibliography}
\end{document}